# Thermosynthesis as Energy source for the RNA World:
## A New Model for the Origin of Life

Anthonie W. J. Muller
Department of Geology, Washington State University, Pullman WA 99164-2812, USA
e-mail: awjmuller@wsu.edu, tel: 509-335-1501, fax: 509-335-7816

**Abstract:** The thermosynthesis concept, biological free energy gain from thermal cycling, is combined with the concept of the RNA World. The resulting overall origin of life model gives new explanations for the emergence of the genetic code and the ribosome. The first protein named $pF_1$ obtains the energy to support the RNA world by a thermal variation of $F_1$ ATP synthase's binding change mechanism. This $pF_1$ is the single translation product during the emergence of the genetic machinery. During thermal cycling $pF_1$ condenses many substrates with broad specificity, yielding NTPs and randomly constituted protein and RNA libraries that contain (self)-replicating RNA. The smallness of $pF_1$ permits the emergence of the genetic machinery by selection of RNA that increases the fraction of $pF_1$s in the protein library: (1) a progenitor of rRNA that concatenates amino acids bound to (2) a chain of 'positional tRNAs' linked by mutual recognition, yielding a $pF_1$ (or its main motif); this positional tRNA set gradually evolves to a set of regular tRNAs functioning according to the genetic code, with concomitant emergence of (3) an mRNA coding for $pF_1$.

Key words: Binding change mechanism—Convection—Dissipative structure—Genetic code—Molecular heat engine—Origin of life — Ribosome—RNA—RNA world— Thermosynthesis

## 1. Introduction

Today's organisms use three types of energy sources: fermentation, photosynthesis and respiration, all complex processes. None of these energy sources have been linked directly to the origin of life. A fourth energy source, 'thermosynthesis,' free energy gain from thermal cycling, has been proposed in a theoretical model for the emergence of the chemiosmotic machinery used by both photosynthesis and respiration (Muller, 1985, 1993, 1995, 1996, 2003). Molecular heat engines produce the same ATP as contemporary ATP synthase, but with much less power because the enzyme turnover time equals the long thermal cycle time of a convection cell (Fig. 1).

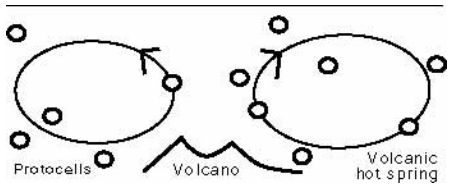

Figure 1. Convection cells in a volcanic hot spring. Protocells carried along by the current are thermally cycled.

The latter constitutes the inanimate self-organizing dissipative structure needed in any origin of life model. According to rRNA sequences, the niche of the last common ancestor of all living organisms was a —plausibly convecting—hot spring (Woese, 1987).
Figure 2 pictures the emergence of the chemiosmotic machinery active during bacterial photosynthesis. The smallest thermosynthesis machine consists of a single protein named $pF_1$. The machine was already previously applied to the protein world (Muller, 1995), and is applied here to the RNA World. $pF_1$ is the progenitor of the β-subunit of the $F_1$ moiety of contemporary membrane-bound $F_oF_1$ ATP synthase (Abrahams *et al.*, 1994). In the binding change mechanism $F_1$ binds ADP and phosphate in the dry enzymatic cleft and forms tightly bound ATP without any energy input and without involvement of covalent intermediates (Boyer, 1993). This ATP is released upon a relay of conformational energy from the $F_o$ part of the enzyme (Fig. 3), which in

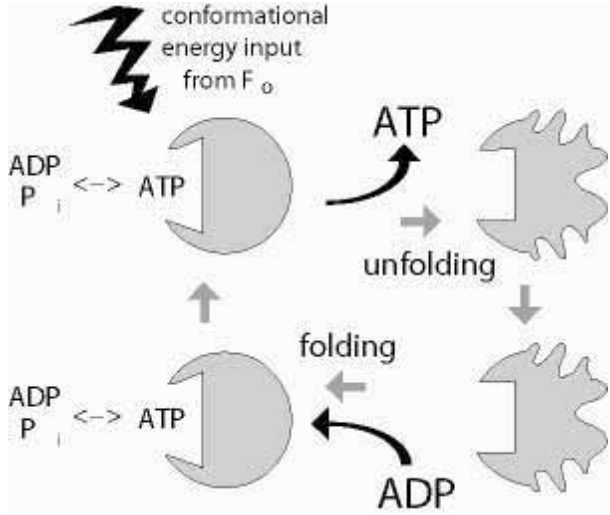

Figure 3. In the binding change mechanism of $F_1$-ATP synthase, tightly bound ATP is spontaneously formed from tightly bound ADP and phosphate. This tightly bound ATP is released upon a conformational energy transfer from the $F_o$ part of the enzyme.

turn obtains the energy from ion transfer across the charged membrane. In the postulated $pF_1$ molecular heat engine, similarly formed and bound ATP is released by thermal unfolding (Fig. 4)

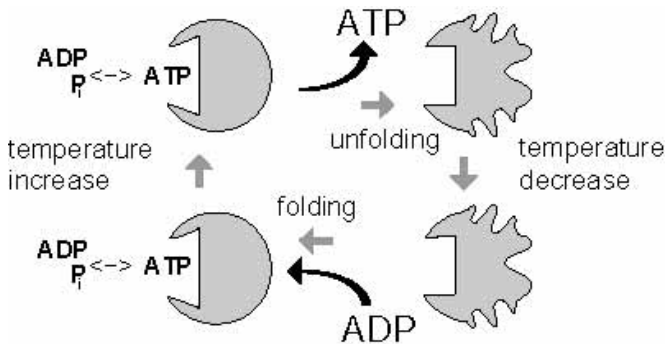

Figure 4. The thermal cycle of convection synchronized with the enzyme cycle of $pF_1$ in the "temperature-induced binding change mechanism." Just as the $F_1$ moiety of today's ATP synthase, the proposed $pF_1$ enzyme can bind ADP and phosphate (bottom); when bound, these substrates are in equilibrium with strongly bound ATP (left). In $pF_1$ this strongly bound ATP is released by a thermal unfolding at high temperature (top). The unfolding of the protein at high temperature takes up heat, the folding at the low temperature releases it. Similar condensation reactions driven by thermal-cycling are proposed for other substrates: the reactions include phosphorylations and the synthesis of peptide bonds.

(Muller, 1995). For comparison, the $\Delta H$ of unfolding of the β-subunit of $F_1$ is ~660 kJ/mole (Wang *et al.*, 1993; Villaverde *et al.*, 1998); the unfolding temperature ($T$) of ~330°K increases ~7,5°K ($\Delta T$) upon nucleotide binding, permitting a Carnot work of $\Delta H (\Delta T / T) = 15$ kJ/mole, the right magnitude for a phosphodiester or peptide bond.
Only in water do proteins and nucleic acids have a higher free energy than their constituent monomers. Under dry conditions the free energies are equal (DeMeis, 1989; Muller, 1995). We apply the rule of parsimony (Benner *et al.*, 1989) to primordial enzymes (Black, 1970) and to enzyme mechanisms: In addition to broadly specific phosphorylations—that yield NMPs, NDPs, NTPs and phospholipids—$pF_1$ can also condense amino acids and peptides to new peptide bonds. In this way, thermosynthesis effects the endergonic synthesis of high-energy products, which considerably simplifies the modeling of primordial metabolism. The basic primordial energy generating mechanism is therefore proposed to be the binding of a substrate in a dehydrated local environment, followed by its conversion into a product with similar free energy in that environment, but a higher free energy in water. The higher free energy makes direct release impossible; this release requires a temperature change.
The research on primordial protein synthesis (Rode, 1999) has not sufficiently advanced to avoid having to postulate the synthesis of long primordial polypeptides. A very small fraction of synthesized *random* protein sequences has $pF_1$ capabilities, associated with a short residue motif. Increasingly *specific* protein sequences have their origin in protein synthesis by ribozymes (Kumar and Yarus, 2001).

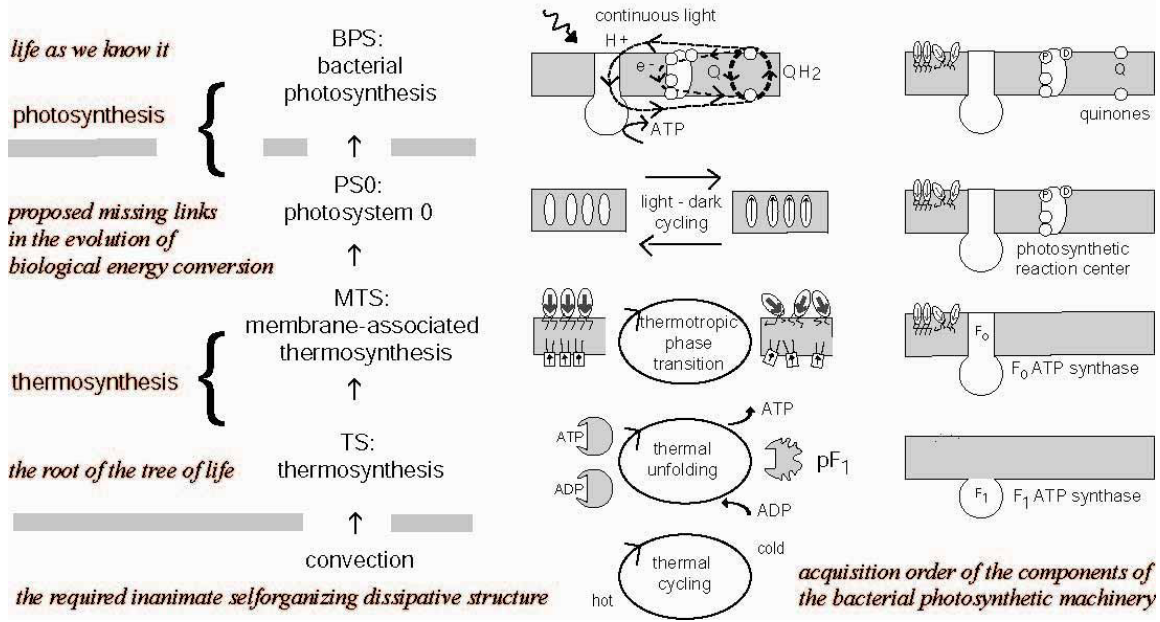

Figure 2. The TS Root and Tree of Life: an evolutionary model for the acquisition of the chemiosmotic machinery as it functions during bacterial photosynthesis (BPS) (Muller, 2003). In BPS, incoming light excites an electron in the reaction center. The electron crosses the membrane along a series of stepping stones where it reduces a quinone, which also picks up a proton from the medium. The resulting quinol diffuses across the membrane where it is oxidized. The electron returns to its origin in the reaction center, and a pumped proton is released to the medium. This proton returns by ATP synthase while ATP is synthesized. The proposed sequence of acquisition is: (1) the $F_1$ part of ATP synthase during the emergence of thermosynthesis (TS, which works on thermal cycling; see Fig. 4); (2) the $F_o$ part of ATP synthase and the lipids of an asymmetric membrane during the emergence of membrane-associated thermosynthesis (MTS, which also works on thermal cycling; the change in membrane potential during a change in membrane dipole potential caused by the temperature drives ATP synthesis); (3) the reaction center with stepping stones for the excited electron during the emergence of Photosystem 0 (PS0, which works on light–dark cycling; here the change in membrane potential due to light-induced dipoles in the photosynthetic reaction centers drives ATP synthesis); and (4) membrane-diffusible quinones during the emergence of BPS, which works in continuous light.

## 2. Emergence of the genetic machinery

The self-organization of life must have involved a self-organizing dissipative structure (Nicolis and Prigogine, 1977; Haken, 1978) that was inanimate. Few such structures are known (Anderson and Stein, 1983). The convection cell is the most prevalent, and, therefore, is a suitable candidate for the origin of life. As required by thermosynthesis, its contents are thermally cycled.

The model presumes the presence of the precursors of protein and RNA synthesis. Several methods have been reported for polypeptide synthesis from amino acids, use of concentrated salt solutions (Rode, 1999) and of mineral and oxide surfaces (White and Erickson, 1981; Basiuk and Sainz-Rojas, 2001). Polymerization on mineral surfaces is energetically favorable, but is slow and leads to oligomers that are strongly bound to the mineral (Orgel, 1998); debonding by a temperature change (Luther *et al.*, 1998) is however plausible.

The genetic machinery emerges in 7 stages (Fig. 5). Protocells, a suitable starting point for the origin of life (Morowitz *et al.*, 1988), are stabilized by membrane lipid phosphorylation by $pF_1$ in stage 1. There are many protocell candidates, some composed of lipids or material found in meteorites (Mautner *et al.*, 1995; Dworkin *et al.*, 2001; Hanczyc *et al.*, 2003; Chen *et al.*, 2004).

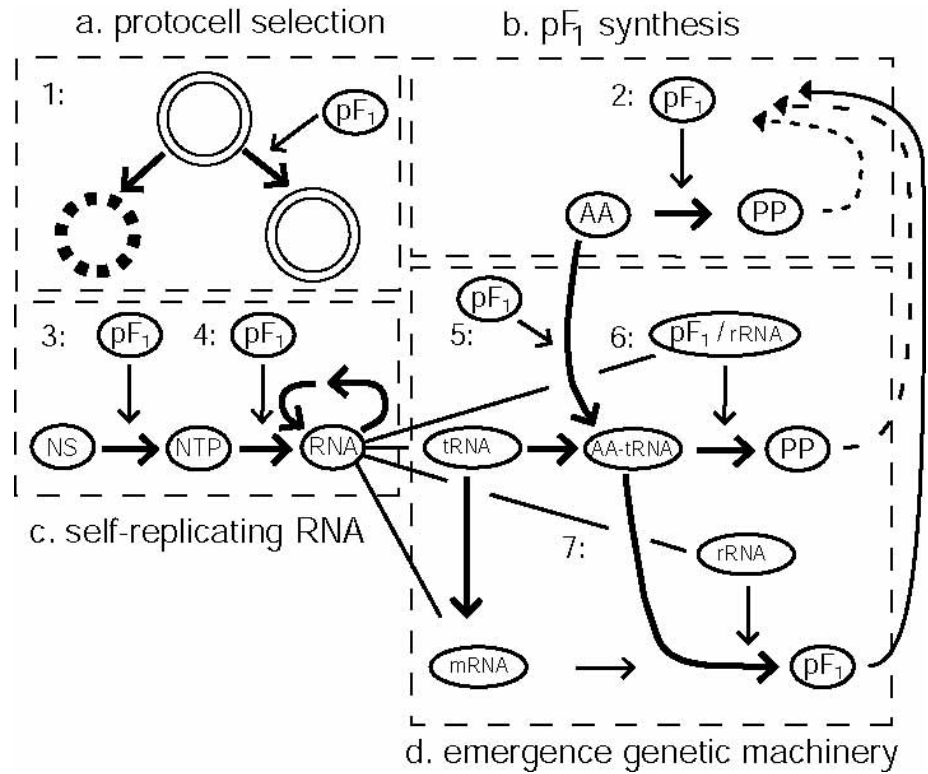

Figure 5. The emergence of the genetic machinery in seven stages. **a** *Protocell selection.* 1. $pF_1$ stabilizes a protocell by membrane lipid phosphorylation. **b** *$pF_1$ synthesis.* 2. in the synthesized library of proteins a small fraction function as $pF_1$: functional $pF_1$ propagation. **c** *Self-replicating RNA.* 3. Nucleotide triphosphates (NTPs) are synthesized from nucleosides (NS). 4. Synthesis of RNA, including self-replicating RNA. From NTPs, $pF_1$ synthesizes RNA both without a template and by RNA copying. RNA is selected that increases the synthesis of $pF_1$. **d** *Emergence genetic machinery.* 5. tRNA is charged with amino acids by ribozymes, or self-charged. 6. tRNA assists in $pF_1$ synthesis. $pF_1$ or rRNA may be present. 7. mRNA emerges from tRNA (see Fig. 6); rRNA, mRNA and tRNAs yield the genetic machinery.

In stage 2, an early $pF_1$ synthesizes by thermosynthesis a library of proteins of which a tiny fraction has multiple substrate condensing ability. In this sense, $pF_1$ propagates functionally, making daughters with similar capability but not necessarily identical composition. Such compositional replication is implausible for proteins (Orgel, 1987): a few small proteins cannot be expected to recognize and copy during peptide bond synthesis the many possible different combinations of amino acid residues. Random synthesis of a specific long protein sequence is also implausible (Orgel, 1987). The $pF_1$ protein must have a short motif sequence that is frequent in a long random sequence. The amino acid residue motif contains only a dehydration pocket and glycine hinges enabling a lobe to cover a substrate in the dehydrated pocket. The lobe resembles the lobe of ATP-using enzymes that consists of the $G(X)_4KT/S(X)_6I/V$ motif (Walker et al., 1982).

The NTPs generated in stage 3 are used for RNA synthesis (Joyce and Orgel, 1993). The self-replicating RNA-replicase of stage 4 is the key theoretical entity of the RNA world. A version that can replicate up to 14 nt has been found (Johnston et al., 2001). RNA replication could resemble the DNA amplification by PCR demonstrated in a convection cell (Krishnan et al., 2002). RNA that enhances the synthesis rate of $pF_1$ is selected.

The emergence of ribozymes with aminoacylation ability constitutes stage 5. Predicted already in 1958 (Crick, 1958), these ribozymes were found in the 90's (Illangasekare et al., 1995; Illangasekare and Yarus, 1999; Yarus and Illangasekare, 1999; Lee et al., 2000; Schimmel and Kelley, 2000; Saito et al., 2001). In stage 6, charged tRNAs increase the overall, still random, protein synthesis rate by their enhanced reactivity (Fig. 6a), catalysed by $pF_1$ or ribozymes (Zhang and Cech, 1997); such ribozymes are progenitors of rRNA.

For our purposes, amino acid activation by RNA (Kumar and Yarus, 2001) is considered to be redundant: no use is made of high energy aminoacyl-AMPs. If this assumption is invalid, aminoacyl-AMP synthesis by $pF_1$ will have to be added to the model.

The first tRNAs contain the amino acid acceptor stem (Weiner and Maizels, 1987). Interaction between tRNAs based on mutual codon:anticodon recognition was considered previously (Crick et al., 1976). Here, modified microhelix tRNAs recognise each other. Not as earlier proposed according to cognate amino acid (Schimmel and Henderson, 1994), but according to position. They recognise a predecessor and are recognised by a follower (Fig. 6b). The 'positional tRNAs' form a tRNA motif sequence

$tRNA_1 tRNA_2 tRNA_3 tRNA_4 tRNA_5 tRNA_6 tRNA_7 \ldots$

coding for the key small residue motif of $pF_1$,

$A_1 A_2 A_3 A_4 A_5 A_6 A_7 \ldots$

The positional aspect is emphasized: although, for example, both $tRNA_3$ and $tRNA_7$ may charge glycine, these two tRNAs can be completely different. Due to the limited recognition length on the tRNA arms, the scheme can only yield a small motif. Assuming four bases to choose from, and a recognition length on the arm of say 3 nt, we obtain a maximal motif sequence length of $4^3 + 1 = 65$. The motif length obviously rapidly increases with recognition length.

The genetic code emerges in stage 7 (Fig. 5). The first positional tRNA of the tRNA motif is the ancestor of today's starting $tRNA_i^{Met}$. From this first positional tRNA an extension emerges to which all positional tRNAs can bind by codon:anticodon interaction (Fig. 6c). This extension resembles the leader in contemporay tRNA precursors that is removed by RNase P (Altman, 1989). The genetic code emerges gradually as the positional tRNAs acquire an anticodon that recognizes a codon on the extension. Mutual tRNA recognition becomes redundant and positional tRNAs evolve to regular tRNAs that also bind to the extension at other codon occurrences (Fig. 6d). A STOP codon is acquired. The advantage of the genetic code is smaller size: one amino acid is coded by only a single codon instead of a larger tRNA. A large set of positional tRNAs is replaced by a smaller set of regular tRNAs that remains small even after its stepwise expansion (Crick, 1968). A break of the extension yields the first template mRNA from tRNA, an ancestry previously proposed (Eigen and Winkler-Oswatitsch, 1981; Maizels and Weiner, 1987). The ribozymes involved in peptide bond synthesis evolve into rRNA, and acquire the ability to act as a ribosome, defined as the particle that translates an mRNA into a protein while translocating along the mRNA (Fig. 6f). In today's large genomes varying the tRNA set could be 'highly disadvantageous' (Crick, 1968) but for a small translation product the effects of mutation in mRNA and tRNA are similar.  Optimization of all RNAs eventually results in a 'frozen' tRNA set (Crick, 1968).  Evolution can then run its course: more mRNA genes and proteins are acquired.  The eventual acquisition of photosynthesis puts an end to the thermal cycling requirement, and $pF_1$ becomes redundant. The transition to isothermy may have been effected by a stepwise transfer of proteins from a thermal cycling dependent operon — ancestor of the heat shock operon (Morita *et al.*, 1999) — to a not thermal-cycling dependent operon. Self-replicating RNA is made redundant by the high fidelity of a DNA based genetic machinery.

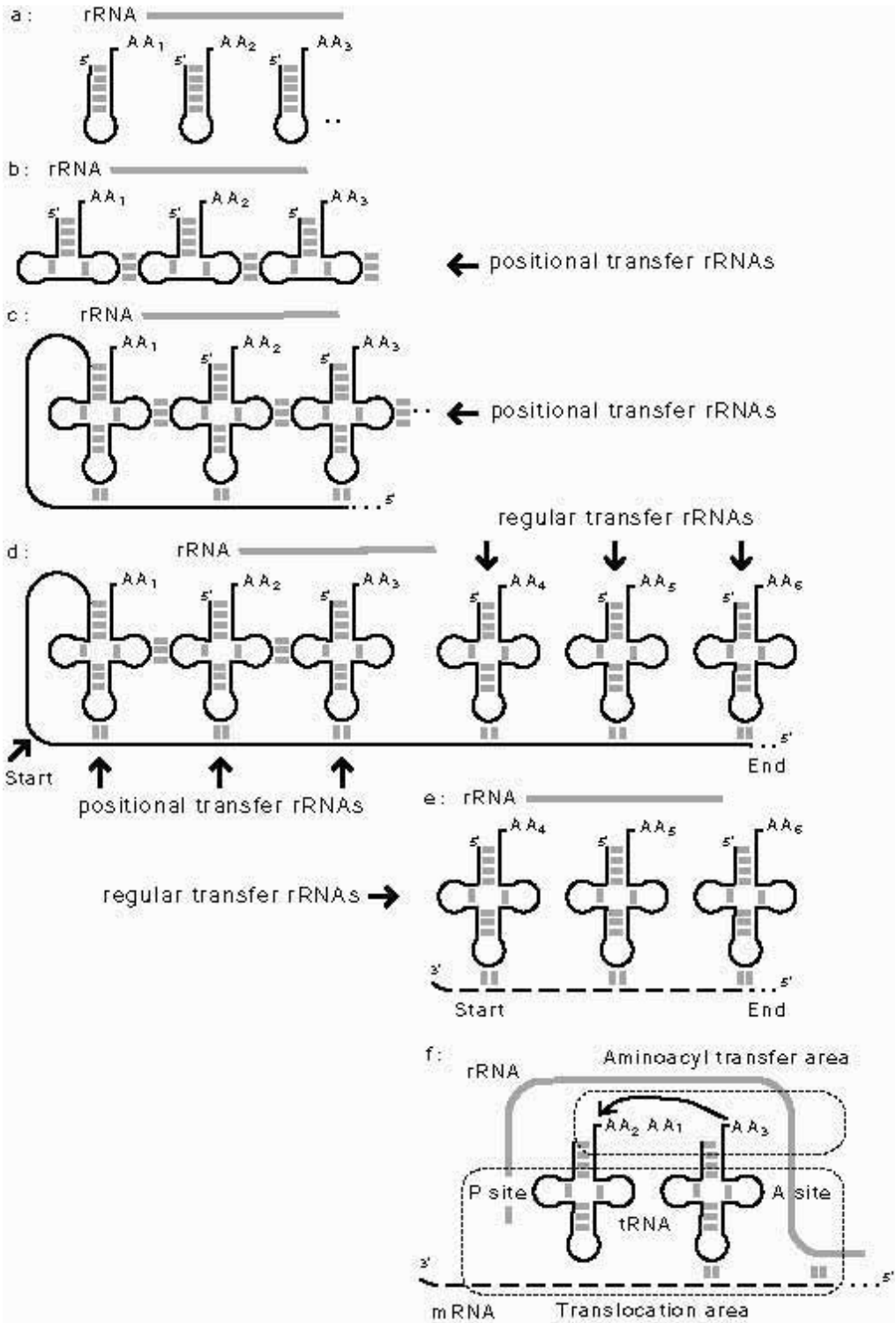

Figure 6. The evolution of tRNA and the emergence of mRNA.  **a** The first tRNAs are minihelices that connect to amino acids yielding charged tRNAs. The charged tRNAs form a protein with random composition, a process catalyzed by an rRNA progenitor or possibly a $pF_1$. **b** Using their arms positional tRNAs recognize a predecessor and a follower, resulting in a tRNA sequence that yields a small protein sequence that is a key motif of $pF_1$. **c** The first tRNA of the sequence extends to a template that binds all other motif tRNAs.  The positional tRNAs acquire anticodons that bind to codons on the extension. **d** The extension is disconnected from the first tRNA (arrow), and yields a mRNA (start at 3' end). Mutual tRNA recognition disappears but tRNA:mRNA recognition remains:  This transition from mutually recognizing positional tRNAs to a set of regular mRNA-codon recognizing tRNAs constitutes the emergence of the genetic code.  **e** Mutual tRNA recognition has disappeared   **f** A ribosome particle containing rRNA emerges that performs (1) aminoacyl transfer from tRNAs in the sequence given by mRNA to a growing peptide chain, and (2) translocates along mRNA in the 3' to 5' direction.

### 3. The Ribosome

During PCR, thermal cycling synchronizes bondings and debondings (Saiki *et al.*, 1985) and it has been applied similarly in a model for the emergence of the ribosome from self-replicating RNA (Campbell, 1991). Temperature effects on translation such as heat shock are in general due to an effect on transcription (Morita *et al.*, 1999; Johansson *et al.*, 2002; Narberhaus 2002; Chowdhury *et al.*, 2003), but several non-trivial temperature effects on the ribosome have been reported as well (Rheinberger and Nierhaus, 1986; Bilgin and Ehrenberg, 1995; Bayfield *et al.*, 2001). The more complex translocation (Joseph, 2003) may have emerged separately from aminoacyl transfer, and may be younger.  Brownian ratchets permit unidirectional movement during external fluctuations, including thermal cycling (Bier 1997; Astumian, 1997), on which progenitors of translocation ratchets (Peskin *et al.,* 1993; Wintermeyer *et al.*, 2004) may have plausibly worked. The origin of the complexity of today's ribosome may therefore result from the need of having to do isothermally what earlier easily could be done by thermal cycling.
We cannot indicate in Fig. 5 where ribozymes replace $pF_1$ during the evolution of protein synthesis.  Many have argued that the ribosome is a ribozyme because only RNA is present at the aminoacyl transfer site, but the ribosome is larger and has little specificity, forming ester, thioester, thioamide and phosphino-amide bonds (Lim and Spirin, 1986). Activation enthalpies and entropies suggest that instead of acting as a specific catalyst, the ribosome during aminoacyl transfer acts as a general catalyst with a dehydrated active site (Sievers *et al*., 2004; Wintermeyer *et al*., 2004) as proposed for $pF_1$ (Muller, 1995): the ribosome may mimic $pF_1$.

**4. Discussion**

The possibility of at least a theoretical solution to the problem of the origin of life has been doubted and the need for an overall model or concept has been stressed (Woese 1980; Medawar and Medawar 1985). The emergence of the genetic machinery especially is considered hard to explain (Trevors and Abel, 2004). The presented model accounts for the self-organization, the thermodynamics, and the emergence of the genetic machinery including the genetic code. The key remaining unsolved issue, the primordial synthesis of a protein library that contains $pF_1$s with the proposed general substrate condensing ability, will have to be solved by experiment.

These condensations during thermal cycling constitute a minimal metabolism that effects the key biochemical reactions, the synthesis of biopolymers. An origin of life model becomes more plausible the smaller its components and the fewer their number. Keeping the energy-converting translation target $pF_1$ short by involving a motif sequence permits a positionally-coding tRNA set to function and to evolve into a regular tRNA set. Smallness also helps by lessening interference due to infidelity during gene replication and protein translation.

Other implications of thermosynthesis were discussed extensively previously (Muller, 1995). In addition we refer to recent studies on nucleic acid polymerisation by convection (Krishnan *et al.*, 2002; Braun *et al.*, 2003; Braun and Libchaber, 2003; Braun, 2004). We identify in $pF_1$ a plausible component of the pre-RNA world (Orgel, 2003), describe this proposed pre-RNA world (Crick, 1993; Dworkin *et al.*, 2003) and illustrate how a protein could have supported the RNA world (Schuster, 1993) by yielding the free energy (Mehta, 1986; Jeffares *et al.*, 1995). We agree with the suggestion that only a few proteins were present during the emergence of the genetic code (Trevors and Abel, 2004).

The free energy gain from heat during thermosynthesis is similar to the synthesis of small peptides on clay by combined wet-dry/thermal cycling: "The free energy needed to drive the condensation reactions is apparently provided by the dehydrating action of the environmental cycles" and not by "high-energy reagents" (White and Erickson, 1980). In contrast with the statement by Rao *et al.* (1980), such clay reactions may therefore agree with the principle of continuity during evolutionary stages (Orgel, 1968).

The direct experimental evidence supporting the thermosynthesis model is: (1) the binding change mechanism of today's ATP synthase, and (2) the ubiquitous role of thermal cycling in germination, propagation and cell division (Muller, 1985). This role during the start of the life cycle of the individual is clearly a possible relic of the origin of life. A chemical reaction involving one turnover during one thermal cycle and supported by a tiny fraction of the proteins present in a library seems detectable by radioisotope methods. The model is testable. Modeling the origin of life has been difficult and the model has little competition. Correctness would explain the elusiveness of the search for the origin of life: a key pertinent process, free energy gain from thermal cycling, would have been overlooked.

**Conclusion**

A generalization of today's ATP synthase mechanism to a primordial thermally dependent process, thermosynthesis, leads in combination with the RNA world to a simple model for the origin of life, including the emergence of the genetic code. The first biological energy source, first metabolism, and first translation product are described. The latter is identified as the direct ancestor of today's ATP synthase. A main remaining issue, the primordial synthesis of proteins, will have to be resolved by experiment. The model integrates many proposals, observations and experimental results. Molecular heat engines may very well have been the main actors during the origin of life.

**Acknowledgements**
I thank Michael Kaufmann, Dirk Schulze-Makuch and Peter Huber for their comments on the manuscript and Adelina Hristova and Lisa Morris for proofreading.